# Room temperature ferromagnetism and giant permittivity in chemical routed $Co_{1.5}Fe_{1.5}O_4$ ferrite particles and their composite with $NaNO_3$

R.N. Bhowmik[*], P. Lokeswara Rao, and J. Udaya Bhanu

Department of Physics, Pondicherry University, R. Venkataraman Nagar, Kalapet, Pondicherry-60014, India.

[*]Author for correspondence (RNB): Tel.:+91-9944064547; Fax: +91-4132655734.

E-mail: rnbhowmik.phy@pondiuni.edu.in

**Abstract**

We report structural, magnetic and dielectric properties of $Co_{1.5}Fe_{1.5}O_4$ nanoparticles and their composites with non-magnetic $NaNO_3$. The samples were derived from metal nitrates solution at different pH values. The chemical routed sample was air heated at 200 $^0$C and 500 $^0$C. Heating of the material showed unusual decrease of crystallite size, but cubic spinel structure is seen in all samples. The samples of $Co_{1.5}Fe_{1.5}O_4$ showed substantially large room temperature ferromagnetic moment, electrical conductivity, dielectric constant, and low dielectric loss. The samples are soft ferromagnet and electrically highly polarized. The interfaces of grains and grain boundaries are actively participating to determine the magnetic and dielectric properties of the ferrite grains. The effects of interfacial contribution are better realized using the ferrite and $NaNO_3$ composite samples. We have examined different scopes of modifying the magnetic and dielectric parameters using same material in pure and composite form.

## 1. INTRODUCTION

Magnetic nanocomposite [1, 2] is one of the exciting topics in present generation research. This is due to the flexibility of modifying magnetic, electric, dielectric and optical properties. In magnetic composites, the magnetic particles are usually added either in an insulating or conductive matrix [3]. Among the different types of magnetic particles, spinel ferrites are well stabilized in different non-magnetic environment and highly promising for the applications in magnetic data storage, magnetic resonance imaging, magnetically guided drug delivery, etc [4, 5]. Spinel ferrites can also be used in microwave devices due to the exhibition of high electrical resistivity and low dielectric loss [6, 7]. Among the spinel ferrites, $Co_xFe_{3-x}O_4$ series ($0 \leq x \leq 3$) is promising candidate for producing multifunctional materials by varying synthesis condition, cobalt content, and particle size and shape dependent parameters (magnetic moment, coercivity, magnetic blocking temperature, conductivity and dielectric constant) [8-10]. $CoFe_2O_4$ is widely studied as a well–known hard magnetic material with large coercivity and magnetization [4, 11, 12]. However, experimental work on Co rich compositions of this series is limited, despite indicating improved magnetic and microwave properties [13], and useful applications in chemical sensors [14], catalytic activity [15], photo-conductive devices [16, 17].

In this work we focus on the $Co_{1.5}Fe_{1.5}O_4$ ferrite. The ferrite nanoparticles were synthesized by co-precipitation of metal nitrates solution at different pH values. The as prepared sample was heated up to 500 $^0$C to study the properties of the material with low annealing temperature side, and study of low temperature annealed samples is few due to the fact that chemical heterogeneity may exist at the microscopic scale of crystal structure. We also prepared a simple magnetic composite of $Co_{1.5}Fe_{1.5}O_4$ particles with $NaNO_3$. The novelty of this magnetic composite is that $NaNO_3$ was naturally formed as the bye-product and coexisting with ferrite particles.

## 2. EXPERIMENTAL

### A. Sample preparation

Fig. 1 shows the schematic presentation of the material synthesis. Stoichiometric amounts of high pure (> 99.999%) $Co(NO_3)_2.6H_2O$ and $Fe(NO_3)_3.9H_2O$ salts were taken to synthesize $Co_{1.5}Fe_{1.5}O_4$. The salts were mixed in distilled water and stirred to prepare a stock solution ($S_0$) of pH value ~ 0.5. The $S_0$ solution was divided into three parts. NaOH solution with pH ~ 13.7 was added separately as precipitating agent in the different parts of the $S_0$ solution in stirring condition and final pH values of the mixed solution were maintained at 12, 11 and 9.5, respectively. Each solution was then heated at 100-110°C with continuous stirring for 2 hours, followed by the heating of the solution at 140-150°C for nearly 1 hour. We observed a black coloured gel at the centre of the beaker and white powder on the wall. X-ray diffraction (XRD) pattern identified this white powder as $NaNO_3$. Each of these dried powders at 140-150°C was divided into two parts (P series and S series). To get pure form of ferrite particles (P series), the bye-product (white coloured $NaNO_3$ powder) was carefully removed from the wall and remaining black coloured gel was washed with distilled water and heated at 110°C. The cleaning process was repeated for several times until there was no more white powder formed on the inner surface of the beaker. Pure form of the black coloured $Co_{1.5}Fe_{1.5}O_4$ samples (P series) obtained from the solutions with pH values 12, 11 and 9.5 were denoted as P1, P2 and P3, respectively. To get the composite material (S series), we did not wash out the white powder from the mixed composite. The mixture of gel and white powder was dried and finally, cooled to room temperature. The mixed powder was ground for 2 hours to make it more homogeneous. The magnetic composite ($Co_{1.5}Fe_{1.5}O_4$ mixed with $NaNO_3$) samples (S batches) obtained from pH values 12, 11 and 9.5 were denoted as S1, S2 and S3, respectively. The samples of P and S batches were then heated at 200 $^0$C for 2 hours and corresponding samples were denoted as P1_200, P2_200, P3_200 for P series and S1_200, S2_200, S3_200 for S series, respectively. The 200 $^0$C samples of P and S series were heated at 500 $^0$C for 2 hours and corresponding samples were denoted as P1_500, P2_500, P3_500 for P series, and S1_500, S2_500, S3_500 for S series, respectively. The rate of heating/cooling was maintained at 5 $^0$C per minute. The heating of the as prepared samples at higher temperature was performed in order to study the thermal effects on crystal structure and physical properties.

**B. Sample characterization and measurements**

The samples were characterized and measured at room temperature. XRD pattern of the samples was recorded using Cu-$K_\alpha$ radiation ($\lambda$ = 1.54056 Å, $2\theta$ range 20 $^0$ to 80 $^0$ with step size 0.02 $^0$ and time/step 2s). Dc magnetization was measured at magnetic field range ± 15 kOe using vibrating sample magnetometer (Lake Shore 7404) and some measurement was carried out using PPMS (Quantum Design, USA). Dielectric properties of the pellet shaped (12 mm diameter, 2-3 mm thickness) samples were measured at 1 Volt ac signal with frequency range 1 Hz to 0.1 MHz using broadband dielectric spectrometer (Novocontrol Techchnology, Germany). The pellets were sandwiched between two gold plated electrodes and measured in a closed sample chamber.

**3. RESULTS AND DISCUSSION**

**A. Structure and morphology**

Fig. 2 shows the profile fit of XRD pattern using FULLPROF program. The profiles confirmed the cubic spinel structure with space group Fd3m. Lattice parameters (8.27-8.28 Å) of the pure form of $Co_{1.5}Fe_{1.5}O_4$ samples are in between the lattice parameters of $CoFe_2O_4$ (~ 8.38Å [11]) and $Co_2FeO_4$ (~ 8.24 Å [13]) particles. Table 1 shows that lattice parameter of the 200 $^0$C samples have increased with the increase of pH value in chemical reaction, where as the lattice parameter of the 500 $^0$C samples has shown decreasing trend with the increase of pH value. Grain/crystallite size has been estimated using the information of prominent XRD peaks (220, 311, 400, 511 and 440) in Debye-Scherrer formula: $<d> = \frac{0.89 \times 180 \times \lambda}{3.14 \times \omega \times \cos\theta_c} nm$. Here, $\theta_c = X_c/2$ and $X_c$ is the peak center in $2\theta$ scale, $\lambda$ = wavelength of X-ray radiation (1.54056 Å), $\omega$ is the Full width at half maximum of the peak counts. Table 1 shows that change of the crystallite size in pure samples is small during the increase of heating temperature from 200 °C to 500 °C. This may be due to less heating time (2 hours). Generally, crystallite size increases with heating temperature, although grain growth process may be slow at low heating temperature [18], due to thermal activated re-crystallization at disordered grain boundary atoms. Interestingly, there is an unusual decrease of crystallite size in pure samples on increasing heating temperature from 200 $^0$C to 500 $^0$C. It shows the crystallites or grain boundary structure of the ferrite nanoparticles is not chemically in proper equilibrium at low temperature heating. The decrease of crystallite size with increasing heating temperature was earlier reported in few materials, which accompanied a lot of surface defects [17] and some cases it indicated a strong magnetic lattice coupling [19]. Unfortunately, most of the reports have studied the nano-structured ferrites with well crystalline structure, which were prepared by chemical routes and heated at higher temperatures to avoid surface defects and structural heterogeneity of the particles. However, properties of the surface defective materials are remarkably advanced from application point of view [20-22]. The crystallite size of the pure samples also shows a non-monotonic increase with increasing pH value, irrespective of heating at 200 °C/500 °C. The samples of P2 (pH value 11) batches showed some exception whose crystallite size is larger than P1 (pH values 12) and P3 (pH values 9.5) batches. Our results of lattice parameter and crystallite size for the samples at different pH values are slightly different in comparison with a monotonic increasing trend with increasing pH values in $CoFe_2O_4$ [23]. In case of composites, the XRD pattern confirmed the coexistence of ferrite and $NaNO_3$. Fig. 3 shows the XRD pattern of S1_200 sample, consisting of P1_200 (ferrite) and $NaNO_3$. XRD data (not shown for all composites) showed the increase of XRD peak counts of $NaNO_3$ component in composite samples with the increase of pH values. XRD profile of $NaNO_3$ is fitted into rhombohedral structure with space group R3C. The novelty of this magnetic composite is that it is naturally formed during the co-precipitated synthesis of ferrites and there is no need to add magnetic ferrite nanoparticles in $NaNO_3$. Such magnetic composite also provide some advantages for studying comparative physical properties, which is one of the objectives of the present work.

**B. Magnetic properties**

Field (H) dependent magnetization (M) of the pure samples at 200 °C is shown in Fig. 4. The M(H) curves (first quadrant only) for 500 °C samples are shown in the inset-a of Fig. 4. All samples in pure form exhibited hysteresis loop, as inset-b of Fig. 4 shows for 200 °C samples. P3_200 sample only showed very small loop, where superparamagnetic features of the particles largely dominate at room temperature. Since crystallite size of all pure samples is below 10 nm, the synthesized samples are consisting of single magnetic domains and smaller size of the lowest pH samples belongs to the superparamagnetic regime [1]. Depending on the strength of the interactions among magnetic domains and magnetic domain structure, P1 and P2 series exhibited strong ferromagnetism at room temperature. Fig. 5 shows a significant decrease of magnetization in the composite samples in comparison with pure samples. Although composite samples showed non-linear increase of M(H) curve and resembling to ferromagnetic feature, Arrot plot (in the inset-a of Fig. 5) using first quadrant of M(H) data does not show any positive intercept of the linear extrapolation of $M^2$ vs. H/M curve from higher field side to the $M^2$ axis only for S3_200 sample. This confirms superparamagnetic behaviour of the ferrite particles in low pH valued samples at room temperature. Rest of the composite samples intercepted the positive side of $M^2$ axis (shown for S3_500 sample in the inset-a of Fig. 5) and showed finite spontaneous ferromagnetic moment. Coercive field ($H_C$) and remanent magnetization ($M_R$) of the samples were calculated from the ferromagnetic loop, as shown in the inset-b of Fig. 5 for S2_500 sample. Table 2 shows the room temperature values of magnetization of each sample at 15 kOe ($M_{15\ kOe}$) and magnetic parameters ($H_C$ and $M_R$). $M_{15\ kOe}$ of the pure samples has increased with heating temperature, irrespective of pH value. However, variation of the $M_{15\ kOe}$ with pH is non-monotonic. For example, $M_{15\ kOe}$ for P2_200 sample is larger in comparison with P1_200 and P3_200 samples. In 500 °C series, $M_{15\ kOe}$ is remarkably large for P3_500 in comparison with P1_500 and P2_500 samples. $M_R$ of the 200 °C samples has decreased with the decrease of pH value, irrespective of the pure and composite samples. In contrast, $M_R$ of the 500 °C pure samples showed significant increase on reducing the pH value in pure sample with minimum value of $M_R$ for P2_500 sample. In composite samples $M_R$ increased with decreasing pH value with a maximum $M_R$ for S2_500 sample. At the same time, the higher pH samples (P1_200 and P1_500) showed larger coercivity ($H_C$), irrespective of the heating temperatures. The result of coercivity at different pH values is different from the reported behavior of $CoFe_2O_4$ particles [23]. Interestingly, lower pH samples have increased their $H_C$ by heating the sample, unlike the decrease of coercivity in higher pH samples. We found that the room temperature magnetization and coercivity of $Co_{1.5}Fe_{1.5}O_4$ ferrite are smaller or comparable to the reported data of $CoFe_2O_4$ particles [11, 18, 20]. The $CoFe_2O_4$ is a hard ferromagnet [24-25], but the present composition $Co_{1.5}Fe_{1.5}O_4$ shows the properties of a good soft ferromagnet at room temperature.

**C. Dielectric Properties**

We examined certain dielectric parameters (ac conductivity ($\sigma$), dielectric constant ($\varepsilon$), dielectric loss (tan$\delta$) and electrical contribution from grains ($\sigma_g$) and grain boundaries ($\sigma_{gb}$)) of the samples. Fig. 6(a) shows that stock solution ($S_0$) with pH value 0.5 is highly conductive with conductivity ~ $8 \times 10^{-3}$ S/cm at frequency 10 Hz. The conductivity, mostly due to conductive ions, rapidly increased with frequency (f) to attain the value ~ $2 \times 10^{-2}$ S/cm at f = 10 kHz. The frequency activated conductivity ($\sigma(f)$) of the stock solution ($S_0$) then saturated above 10 kHz, which indicated leveling of ionic contributions at higher frequencies and electrical transport between two electrodes may be due to free motions of the ions. $\sigma(f)$ has significantly decreased in the pure form of ferrite particles in comparison with stock solution ($S_0$). The ferrite samples also showed relatively slow $\sigma(f)$ in 200 $^0$C (Fig. 6(a)) and 500 $^0$C (Fig. 6(b)) samples. $\sigma(f)$ of the sample P3_500 is very weak up to 1 kHz, which is followed by a sharp increase of conductivity. Over all conductivity of the 200 $^0$C samples is higher than the 500 $^0$C samples. The general trend is that conductivity of the material has increased with the increase of pH value. The $\sigma(f)$ of the composite samples at 200 $^0$C (Fig. 6(c)) and at 500 $^0$C (Fig. 6(d)) is slow and magnitude wise small in comparison with pure samples. We understand that electrical conductivity of the composite samples at grain boundaries is strongly affected by the coexistence of relatively poor ionic conductor (dc conductivity ~$10^{-9}$ S/cm estimated from Fig. 6(c)) $NaNO_3$ in the composites. The basic property of increasing conductivity with increasing pH value or the decrease of conductivity by increasing the heating temperature from 200 $^0$C to 500 $^0$C of the as prepared samples are also retained in composite samples. The conductivity values at 1 Hz ($\sigma_{1\ Hz}$) of all ferrite samples are shown in Table 3. We noted that pure form of the ferrite samples are sufficiently conductive at room temperature (> $10^{-4}$ S/cm) and could be the potential candidate for solid state fuel applications [26]. As shown in Fig. 6, $\sigma(f)$ of ferrite samples are fitted with Jonscher power law: $\sigma(f) \sim f^n$ with at least two exponent ($n_1$ and $n_2$) values. The exponent values of pure and composite samples at lower frequency regime ($n_1$) and higher frequency regime ($n_2$) are always less than 0.4, except $n_2 \sim 0.70$ for P3_500 sample. This indicated a multiple hopping, mixed with long range hopping, of electronic charge carriers (polarons) at the grains and grain boundaries [27-28]. The $n_1$ values (0.18-0.19) are nearly same for P1_200, P2_200 and P3_200 samples, where as $n_2$ value decreases from 0.33 to 0.20 on decreasing the pH value from 12 to 9.5. This means short ranged hopping of polarons or electrons inside the grains are more favoured for the samples with high pH value. For 500 $^0$C samples, $n_1$ value decreases from 0.33 to 0.05 with the decrease of pH value. This indicated slowing down of the dynamics of bound charge carriers (polarons) as an effect of reduced surface defects in heated samples. On the other hand, increase of $n_2$ values at 500 $^0$C from 0.19 to 0.70 on decreasing the pH value suggests that electronic charge carriers become more localized at the surface of grains and forming short ranged polarons at the interfaces of grains and grain boundaries. The values of $n_1$ and $n_2$ obtained from 500 $^0$C composite samples are significantly small in comparison with pure samples. This is due to coexistence of mobile ions ($Na^+$, $NO_3^-$) from ionic conductor $NaNO_3$ at the surfaces of grains and grain boundaries. It is interesting to note that $n_1$ of the composite samples are significantly higher than $n_2$, unlike a different trend observed in pure samples. This shows that $NaNO_3$ is affecting in the formation of localized polarons at the grain boundaries of the composite samples. In the composites, grains are consisting of highly conducting $Co_{1.5}Fe_{1.5}O_4$ particles (domains). It is believed that grain boundaries of ferrites are more active at lower frequencies. As the frequency increases the grains become more active and short ranged hopping of electrons between $Fe^{2+} \leftrightarrow Fe^{3+}$ and holes between $Co^{2+} \leftrightarrow Co^{3+}$ ions in the octahedral sites of the cubic spinel structure is activated [28-29]. The hopping of electrons between $Fe^{2+} \leftrightarrow Fe^{3+}$ in the octahedral sites is suppressed in the composite samples. As the frequency is increased, the electrons do not follow the frequencies of applied ac electric field. This leads to weakening of the frequency activated conductivity at higher frequencies with small values of $n_2$.

The Cole-Cole plots ($-Z^{//}$ vs. $Z^{/}$) of complex impedance spectrum (Fig. 7(a)) showed no clear signature of a semi-circle at the lower frequency side, i.e., higher values of $Z^{/}$. This reveals a perturbed impedance contribution from grain boundary or defective surface of ferrite particles of 200 $^0$C samples. In the absence of any straight line in Cole-Cole plot of complex impedance, we suggest that the curved line is not due to a typical electrode effect. Rather, the concaved upward curve shows the interfacial effect of the complex microstructure at the grain boundaries [28, 30]. A small semi-circle at higher frequency side, i.e., at smaller values of $Z^{/}$, suggests a well defined impedance contribution mainly from grains. Fig. 7(b) shows that interfacial effect still dominates in 500 $^0$C pure (P1_500 and P2_500) samples. The appearance of two coexisting semi-circles in P3_500 (low pH) sample gives the indication of the separation of grain boundary impedance ($R_{gb}$) at low frequencies from the contribution of grains ($R_g$) at high frequencies. There is a dramatic increase of the $R_g$ in the composite samples at 200 $^0$C (Fig. 7(c)) and 500 $^0$C (Fig. 7(d)) in addition to a strong perturbation in the grain boundary contribution. The magnitudes of grain ($\rho_g$) and grain boundary ($\rho_{gb}$) resistivity were obtained from $R_g$ and $R_{gb}$ of the samples using fitted data of Cole-Cole plot. As in Table 3, the trend of resistivity (inverse of conductivity) is well consistent to the features observed from ac conductivity at 1 Hz.

Fig. 8 indicated the signature of giant relative permittivity/dielectric constant ($\varepsilon$) in pure and composite form of the samples, as well as in the stock solution ($S_0$). The stock solution ($S_0$) is highly polarized under the application of ac field and typical $\varepsilon$ at 1 Hz is ~ $1.616 \times 10^9$. When the ions of the stock solution ($S_0$) become more localized in the cubic spinel structure, the $\varepsilon$ value significantly decreases and can be understood from the relatively lower value of $\varepsilon \sim 10^6\text{-}10^8$ at 1 Hz in pure samples. The dielectric constant is further decreased in composite samples due to the coexistence of weakly polarized molecules of $NaNO_3$ whose permittivity is nearly 3 orders less in comparison with pure ferrite samples. Typical magnitude of $\varepsilon$ at 1 Hz is shown in Table 3 for all samples. It may be noted that magnitude of $\varepsilon$ gradually decreases with the increase of frequency and exhibited a typical value ~ $10^2\text{-}10^5$ at 100 kHz. Such huge values of $\varepsilon$ over a large frequency regime show highly polarized nature of the samples [31]. Electrical polarization in the present ferrite material aroused due to the short ranged displacements of B sites cations ($Fe^{3+}$, $Fe^{2+}$, $Co^{2+}$, $Co^{3+}$) or hopping of charge carriers among the cations (electron hopping: $Fe^{2+} \leftrightarrow Fe^{3+}$, hole hopping: $Co^{3+} \leftrightarrow Co^{2+}$). The effects of site exchange of cations ($Fe^{3+}$, $Co^{2+}$) among A and B sites of the spinel structure may be expected, but this is generally observed for the samples heated at high temperatures ($\leq 800\ ^0C$). However, one can expect a large amount of electrical polarization from the heterogeneous electronic microstructure at the grain boundaries [28, 30]. As the frequency of the ac field increased, the forward and backward motions of the electronic charge carries stored at the interfaces of grains lag behind the driving frequency (f) of ac field. This reduces interfacial polarization of the samples at higher frequencies, but polarization from internal dynamics of the ferrite samples (e.g., short ranged displacements of B sites cations, hopping of charge carriers among the cations, etc) maintains the large dielectric constant ($\varepsilon$) up to higher frequency. Table 3 shows that $\varepsilon$ depend on pH and heating temperature of the samples in pure and composite form. The general tendency is that ε decreased with decreasing pH values, irrespective of heating temperature. Heating of the samples from 200 $^0$C to 500 $^0$C has decreased the magnitude of ε up to 5-10 times depending on the pure and composite form of materials. We attribute the decrease of ε in 500 $^0$C samples to the reduction of polarization originated from heterogeneous electronic microstructure of samples at 200 $^0$C. Fig. 9 (a-d) shows that tanδ of the stock solution ($S_0$) is very small (0.2-2.5) and the value has increased to 1-9 and 1-11 for pure and composite form of samples, respectively. This means tanδ of the ferrite particles in pure samples has not changed significantly in composites. This is due to the fact that tanδ of the ionic conductor $NaNO_3$ (~0.03-2 in (Fig. 9(c)) is not adding much contribution in composite samples. Low dielectric loss of the samples below 100 Hz indirectly suggests that huge dielectric constant is not an artifact of the electrode effects, it may be the intrinsic nature of the samples.

It is interesting to note that pure and composite form of the samples exhibited a tanδ peak at finite frequency ($f_P$), which can be close to the hopping frequency of electrical charge carriers between two successive ionic sites of the cations ($Fe^{2+} \leftrightarrow Fe^{3+}$ and $Co^{3+} \leftrightarrow Co^{2+}$). The tanδ peak at higher frequencies suggests an appreciable contribution of the orientation of electrical dipoles to the dielectric relaxation process mainly at grains. These dipoles originated due to electron density modulation of the multi-valence cation pairs ($Fe^{2+}$ (more electron density) $\leftrightarrow Fe^{3+}$ (less electron density) and $Co^{3+}$ (less electron density) $\leftrightarrow Co^{2+}$ (more electron density)) [32-33]. On increasing the pH, the peak frequency ($f_P$) shifts to lower value in P series of 200 $^0$C samples (Fig. 9(a)), whereas $f_P$ shifts to higher values for 500 $^0$C samples (Fig. 9(b)). Sufficiently large value ($10^{-3}$-$10^{-4}$s) of the relaxation time/hopping time ($\tau_m = 1/f_p$) of the dipoles shows strong interactions among the electronic charge carriers (polarons). The origin of strong interactions may not be electrostatic alone; rather magnetic interactions between B site magnetic spins of the grains are also affecting the dielectric relaxation process. Table 3 shows that $\tau_m$ for the pure form of 500 $^0$C samples increases with the decrease of pH values, unlike a decreasing trend of $\tau_m$ in 200 $^0$C samples. $\tau_m$ of the composites at 500 $^0$C also decreased with the decrease of pH values, but value is relatively small in comparison with pure counter parts. The fast relaxation process in composite samples supports the effect of magnetic interactions in dielectric relaxation process, because non-magnetic $NaNO_3$ helped to reduce the magnetic interactions among ferromagnetic grains of composite samples. The reduced magnetic interactions made the hopping mechanism of charge carriers short ranged type in composite samples, whereas variable long range hopping of the charge carriers are dominating the relaxation dynamics of pure samples.

We understand the properties of the samples annealed at lower temperatures by a schematic model in Fig. 10. We used nitrate salts and NaOH in co-precipitation route, where different types of positive ($Fe^{3+}$ + $Co^{2+}$ +$H^+$+ $Na^+$) and negative ($NO_3^-$ + $OH^-$) ions coexisted in the precursor solution (see Fig. 1). The number of reagent ions ($Na^+$ and $NO_3^-$) is more for the solution at higher pH, because of higher amount of NaOH in a fixed volume of nitrate solution. The reagent ions (or compound $NaNO_3$) were sitting at the surfaces of grains or grain boundaries during the growth of ferrite particles. Fig. 10(a) suggests that magnetic spins and bye-products (ions or salts) form an inhomogeneous defective layer surrounding the ferromagnetic domains. The effects of such layers are prominent for the samples at higher pH value. When the chemical routed sample is heated the chemical surfactants (bye-products) that retained at the surfaces (grain boundaries) are removed from the particles. The removal of sitting ions introduces crystal defects/porosity at the sitting sites of the defective layers (Fig. 10(b)). There is a possibility that large number of surface spins can be pinned at the defective sites of surfaces. Applied thermal energy during heating of the as prepared samples is spent mainly to reduce the volume of the surfactants induced crystal defects at outer layers (grain boundaries) of the nano-sized crystals, leaving a little (or nearly zero) energy that can activate crystal growth process. Hence, crystallite size of the samples heated at low temperature range 200 $^0$C to 500 $^0$C is effectively reduced due to decrease of surface defects volume (Fig. 10(c)). In the absence of less number of defect sites at 500 $^0$C samples, we suggest that less number of surface spins is pinned (Fig. 10(c)). This results in increasing magnetization with reduced coercivity in 500 $^0$C samples of the P1 and P2 series. The defective surface layer is magnetically more active due to additional surface stress [21, 34] created by the sitting ions. The pinning of magnetic domains at the heterogeneous grain boundaries/surfaces increased coercivity, and showed better room temperature ferromagnetism in the samples with higher pH value. The creation of magnetic nanoparticles with surface spin pinning could be an alternative technique for overcoming the superparamagnetic fluctuation in nano-sized magnetic particles [35]. The samples with less pH value (e.g., P3_200, S3_200) are largely superparamagnetic in nature. Since superparamagnetism is a typical character where magnetic exchange interactions inside the single domains are overcome by the thermal activated random motion of the domains, we suggest that surface spins are magnetically less active for the samples with less pH value. In this case, surface spins are easily allowed for the random freezing on the less defective surfaces. This results in the decrease of ferromagnetic moment and inter-particles interactions. The decrease of magnetic moment in low pH samples is different from the observations in composite samples. The coexistence of non-magnetic $NaNO_3$ reduces effective magnetic volume and inter-particles magnetic interactions in composite samples. There is a probability that electronic charge carriers are highly mobile in the defective surface layers of the samples. This concept is matching to the higher values of electrical conductivity, dielectric constant and relaxation time of the samples prepared at higher pH value and heated at lower temperature (200 $^0$C). The presence of poor conductor NaNO$_3$ reduced the effective conductivity of the ferrite particles in the composite samples. But, the fundamental property of higher conductivity in samples with higher pH is unchanged in the composite samples also and this validates our assumptions in the proposed schematic diagram.

**D. Conclusions**

This experimental work highlighted the magnetic and dielectric properties of nano-structured Co$_{1.5}$Fe$_{1.5}$O$_4$ ferrite, which was synthesized in chemical co-precipitation and subsequent heating at low temperature regime. Crystallite size of the samples is in single domain range. An unusual decrease of crystallite size by heating the co-precipitated sample is attributed to the reduction of defective surface volume. Magnetic and dielectric properties of the pure and composite form of Co$_{1.5}$Fe$_{1.5}$O$_4$ nanoparticles strongly depend on the pH value during coprecipitation and pinning of surface spins at defective sites. Ferromagnetic moment in composite samples is decreased due to the coexistence of non-magnetic NaNO$_3$. Superparamagnetic features dominate in the samples obtained at lower pH value, irrespective of pure or composite form of Co$_{1.5}$Fe$_{1.5}$O$_4$ nanoparticles. The samples obtained at higher pH value are highly soft ferromagnetic with large moment, as well as highly conductive and exhibited large dielectric constant. The results promise a wide scope of varying magnetic and dielectric properties by engineering pH dependent surface defects in pure and composite form of Co$_{1.5}$Fe$_{1.5}$O$_4$ particles. This is useful for room temperature applications of ferrite nanoparticles.

Acknowledgment

We thank to CIF, Pondicherry University for providing experimental facilities and also to Dr. A. Banerjee of UGC-DAE CSR, Indore, for doing magnetic measurement of some samples.

**REFERENCES**

[1] Y.-wook Jun, J.-wook Seo, and J. Cheon, Accounts of Chem. Res. 41 (2008) 179 .

[2] S. Li, M.M. Lin, M.S. Toprak, D.K. Kim, and M. Muhammed, Nano Reviews 1 (2010) 5214.

[3] S.-H. Yu, and M. Yoshimura, Adv. Funct. Mater 1 (2002) 9.

[4] R. Skomski, J. Phys.: Condens. Matter 15 (2003) R841.

[5] A. Sandhu, H. Handa, and M. Abe, Nanotechnology 21 (2010) 442001.

[6] B.K. Kuanr, V.Veerakumar, K. Lingam, S. R. Mishra, A.V. Kuanr, R.E. Camley, and Z. Celinski, J. Appl. Phys. 105 (2009) 07B522.

[7] V.G. Harris, A. Geiler, Y. Chen, S. D. Yoon, M. Wu, A. Yang, Z. Chen, P. He, P. V. Parimi, X. Zuo, C.E. Patton, M. Abe, O. Acher, and C. Vittoria, J. Magn. Magn. Mater. 321 (2009) 2035.

[8] A.R. West, H. Kawai, H. Kageyam, M. Tabuchi, M. Nagata, and H. Tukamoto, J. Mater. Chem. 11 (2001) 1662.

[9] A. C. C. Tseung, and J. R. Goldstein, J. Material Science 7 (1972) 1383.

[10] M.T. Klem, M. Young, and T. Douglas, Materials Today 8 (2005) 28.

[11] Z. Zi, Y. Sun, X. Zhu, Z.Yang, J. Dai, W. Song, J.Magn.Magn. Mater. 321 (2009) 1251.

[12] V. Pallai, and D.O. Shah, J. Magn. Magn. Mater. 163 (1996) 243.

[13] I. P. Muthuselvam, and R.N. Bhowmik, Solid State Sci. 11 (2009) 719.

[14] X. S. Niu, W. P. Du, and W. M. Du, Sens. Actuators B 99 (2004) 405.

[15] L.F. Liotta, and G. D. Carlo, et al., Appl. Catal. B 70 (2007) 314.

[16] A. K. Giri, E. M. Kirkpatrick, P. Moongkhamklang, S. A. Majetich, and V.G. Harris, Appl. Phys. Lett. 80 (2002) 2341.

[17] Z. Zhu, X. Li, Q. Zhao, Y. Shi, H. Li, and G. Chen, J. Nanopart. Res. 13 (2011) 2147.

[18] Vinod Kumar, M. S. Yadav, and R. P. Pant, J. Magn. Magn. Mater. 320 (2008) 1729.

[19] X. G. Zheng, H. Kubozono, H. Yamada, K. Kato, Y. Ishiwata, and C. N. Xu, Nature Nanotechnology 3 (2008) 724.

[20] K. E. Mooney, J. A. Nelson, and M. J. Wagner, Chem. Mater. 16 (2004) 3155.

[21] F. Casoli, L. Nasi, F. Albertini, S. Fabbrici, C. Bocchi, F. Germini, P. Luches, A. Rota, and S. Valeri, J. Appl. Phys. 103 (2008) 043912.

[22] X.P. Zhong, G.H. Wang, and H.L. Luo, J. Magn. Magn. Mater. 120 (1993) 190.

[23] V. Kumar, A. Rana, N. Kumar, and R.P. Pant, Int. J. Appl. Ceram. Technol. 8 (2011) 120.

[24] Y. Cedeño-Mattei, and O. Perales-Pérez, Microelectronics Journal 40 (2009) 673.

[25] Y. Qu, H. Yang, N. Yang, Y. Fan, H. Zhu, and G. Zou, Mater. Lett. 60 (2006) 3548.

[26] W.Z. Zhu, and S.C. Deevi, Mater. Sci. Eng. A 362 (2003) 228.

[27] I. P. Muthuselvam, and R.N. Bhowmik, J. Phys. D: Appl. Phys. 43 (2010) 465002.

[28] M. George, S.S. Nair, K.A. Malini, P.A. Joy, and M.R, Anantharaman, J. Phys. D: Appl. Phys. 40 (2007) 1593.

[29] A.M.M. Farea, S. Kumar, K.M. Batoo, A. Yousef, and Alimuddin, Physica B: Condensed Matter 403 (2008) 684.

[30] R. Martínez, A. Kumar, R. Palai, J.F. Scott, and R.S. Katiyar, J. Phys. D: Appl. Phys. 44 (2011) 105302.

[31] M. Viviani, M. Bassoli, V. Buscaglia, M.T. Buscaglia, and P. Nanni, J. Phys. D: Appl. Phys. 42 (2009) 175407.

[32] M.A. Subramanian, T. He, J. Chen, N.S. Rogado, T.G. Calvarese, and A.W. Sleight, Adv. Mater. 18 (2006) 1737.

[33] R.N. Bhowmik, R. Ranganathan, B. Ghosh, S. Kumar, and S. Chattopadhyay, J. Alloys Compd. 456 (2008) 348.

[34] R. N. Bhowmik, R. Ranganathan, R. Nagarajan, B. Ghosh, and S. Kumar, Phys. Rev. B 72 (2005) 094405.

[35] V. Skumryev, S. Stoyanov, Y. Zhang, G. Hadjipanayis, D. Givord, and J. Nogués, Nature 423 (2003) 850.

Table 1. Lattice parameter (a), Crystallite size (d), cell volume (V) of pure samples and $NaNO_3$.

| Sample | Lattice constant, a (Å) | Cell volume, $a^3$ ($10^{-30}$ $m^3$) | Crystallite size $<d>_{311}$ (nm) | Crystallite size, $<d>_{avg}$ (nm) for 6 prominent peaks |
|---|---|---|---|---|
| P1_200 | 8.293 ± 0.002 | 570.34 ± 0.20 | 9.46 | 8.64 |
| P2_200 | 8.276 ± 0.002 | 566.84 ± 0.25 | 10.1 | 9.08 |
| P3_200 | 8.249 ± 0.003 | 561.39 ± 0.37 | 8.6 | 7.95 |
| P1_500 | 8.257 ± 0.001 | 562.95 ± 0.06 | 9.26 | 8.36 |
| P2_500 | 8.272 ± 0.002 | 566.02 ± 0.19 | 9.8 | 8.71 |
| P3_500 | 8.278 ± 0.002 | 567.25 ± 0.25 | 7.94 | 6.57 |
| $NaNO_3$ | a=b=5.07663 ± 0.00096 c=16.85566 ± 0.00366 | 376.21 ± 0.13 | 62.2 | |

Table 2. List of M at 15 kOe, $M_R$, $H_C$ for samples measured at 300K with different pH values and annealing temperatures.

| Sample | M at 15 kOe (emu/g) | $M_R$ (emu/g) | $H_C$ (Oe) |
|---|---|---|---|
| P1_200 | 16.39 | 1.29 | 77.51 |
| P2_200 | 23.69 | 0.93 | 32.86 |
| P3_200 | 14.58 | 0.05 | 5.05 |
| P1_500 | 29.47 | 1.31 | 35.71 |
| P2_500 | 28.19 | 0.56 | 12.59 |
| P3_500 | 34.05 | 3.15 | 9.42 |
| S1_200 | 8.23 | 0.36 | 37.26 |
| S3_200 | 4.03 | 0.0 | 0.0 |
| S1_500 | 10.32 | 0.25 | 19.26 |
| S2_500 | 8.91 | 0.58 | 58.56 |
| S3_500 | 6.41 | 0.13 | 25.56 |

Table 3. List of $\sigma_{1Hz}$, $\varepsilon_{1Hz}$, resistivity $\rho$, relaxation time $\tau_m$ for samples of different pH values and annealing temperatures.

| Sample | $\sigma_{1Hz}$ ($10^{-6}$ S/cm) | $n_1$ | $n_2$ | $\rho_g$ ($\Omega$-cm) | $\rho_{gb}$ ($\Omega$-cm) | $\varepsilon_{1Hz}$ ($10^6$) | $\tau_m$ ($10^{-4}$ sec) |
|---|---|---|---|---|---|---|---|
| P1_200 | 534.07 | 0.19 | 0.33 | 670 | -- | 960 | 8.33 |
| P2_200 | 106.26 | 0.19 | 0.24 | 1516 | --- | 191 | 3.87 |
| P3_200 | 185.57 | 0.18 | 0.20 | 1562 | -- | 334 | 3.87 |
| P1_500 | 62.38 | 0.33 | 0.19 | 395 | -- | 112 | 5.15 |
| P2_500 | 35.39 | 0.28 | 0.23 | 3250 | -- | 64 | 19.00 |
| P3_500 | 17.21 | 0.05 | 0.70 | 32722 | 144.4M | 31 | 208.00 |
| S1_200 | 80.30 | 0.22 | 0.07 | 3482 | -- | 144 | --- |
| S3_200 | 2.70 | 0.07 | 0.03 | 185105 | 5.52x$10^5$ | 5 | 1.97 |
| S1_500 | 89.34 | 0.38 | 0.14 | 3810 | -- | 49 | 14.22 |
| S2_500 | 4.71 | 0.28 | 0.08 | 26682 | 6.02 x$10^5$ | 8 | 11.24 |
| S3_500 | 3.68 | 0.13 | 0.05 | 89802 | 5.48x$10^5$ | 7 | 1.16 |
| NaNO$_3$ | 1.81×$10^{-3}$ | 0.08 | 0.94 | ---- | 2.14x$10^8$ | 3.26×$10^{-3}$ | 333.33 |

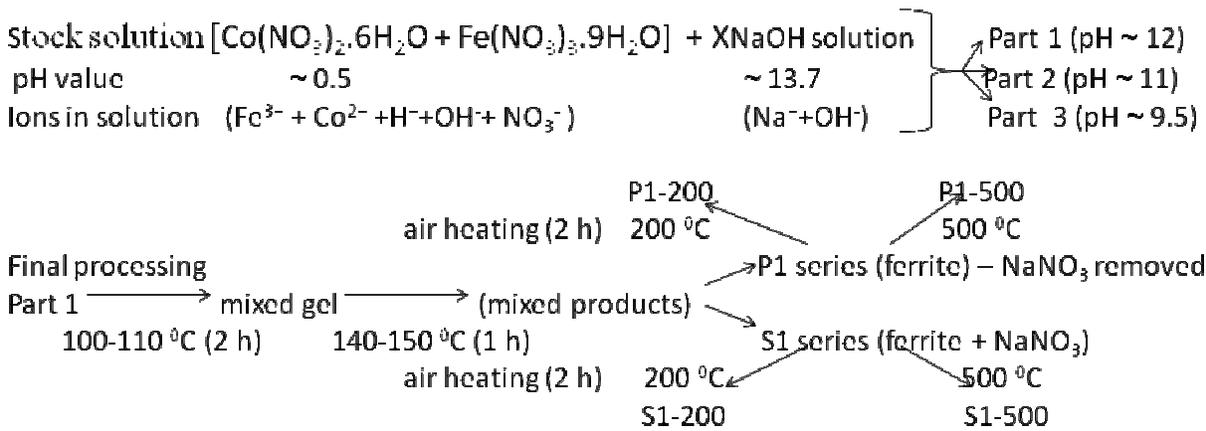

Similar steps, as the case for Part 1, were followed for Part 2 and Part 3 also to get the samples P2 series (P2-200, S2-200, P2-500, S2-500), and P3 series (P2-200, S2-200, P2-500, S2-500), respectively.

Fig. 1 Schematic diagram of the synthesis procedure of the pure and composite form of samples.

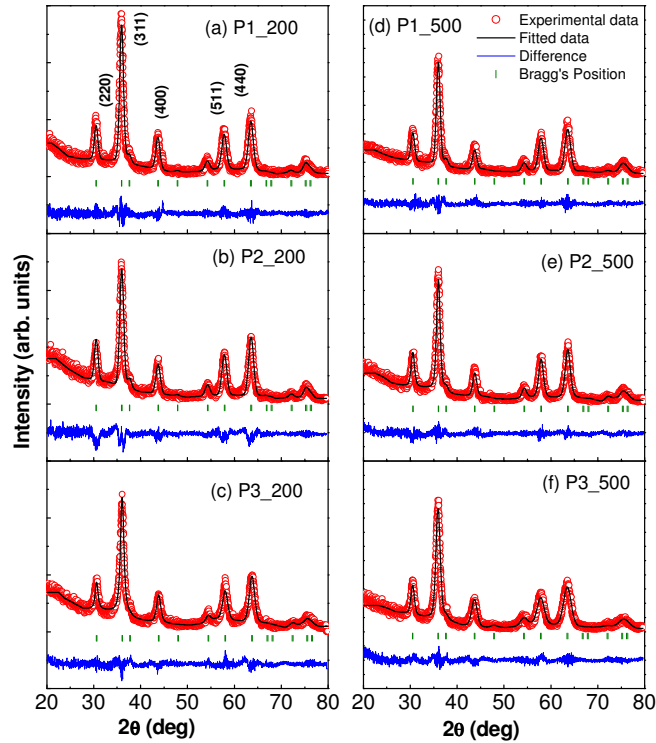

Fig. 2 (Color online) XRD profile fit of pure samples synthesized at different pH values. The as prepared samples were annealed at 200 $^{o}$C (a-c) and 500 $^{o}$C (d-f), respectively.

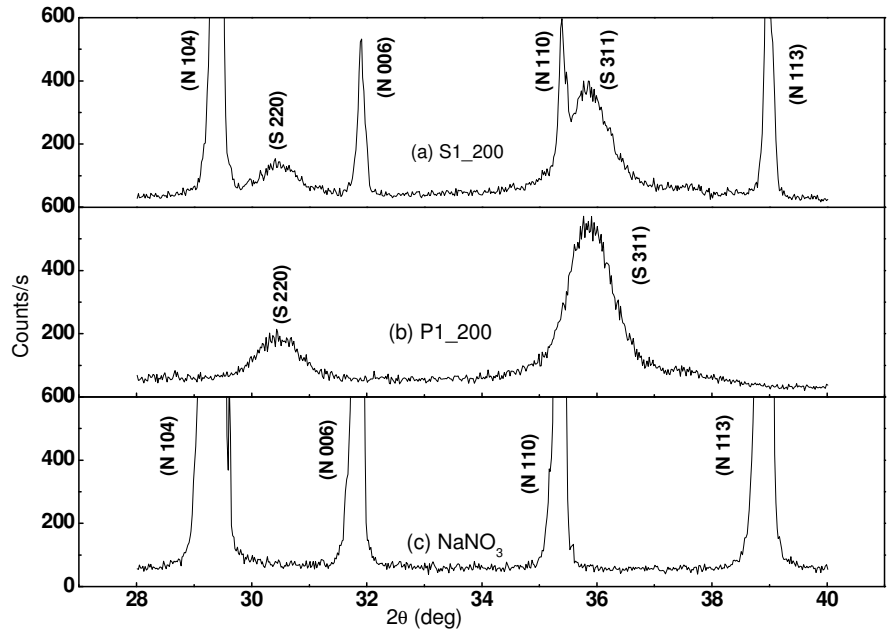

Fig. 3 XRD pattern of S1_200 (a), P1_200 (b), and NaNO$_3$ (c) samples. XRD patterns of the samples have been plotted in the limited range of (x-y) scales for clarity.

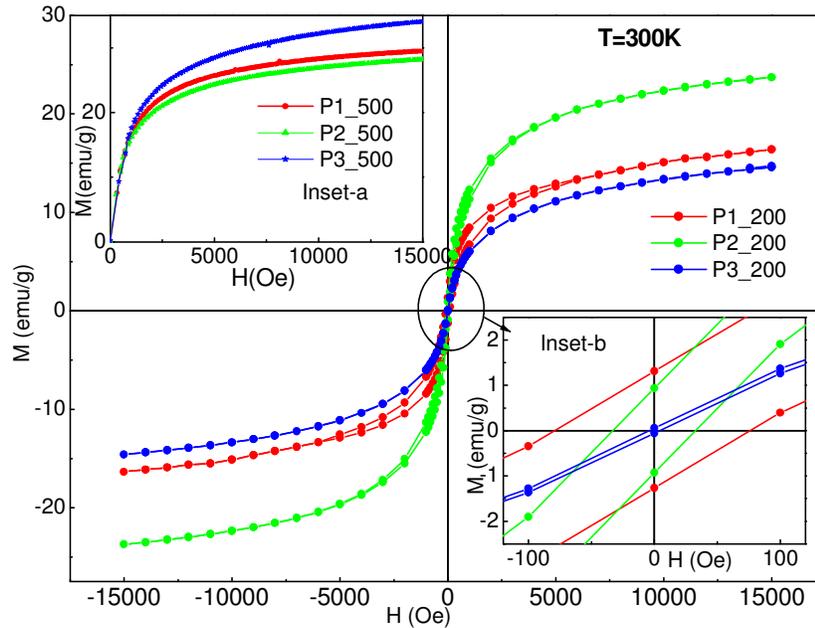

Fig. 4 (Color online) M(H) loops at 300 K for pure samples at 200 $^0$C. Inset-a shows the M(H) data at first quadrant for 500 $^0$C heated pure samples. Inset-b shows the loop of 200 $^0$C samples at lower magnetic field.

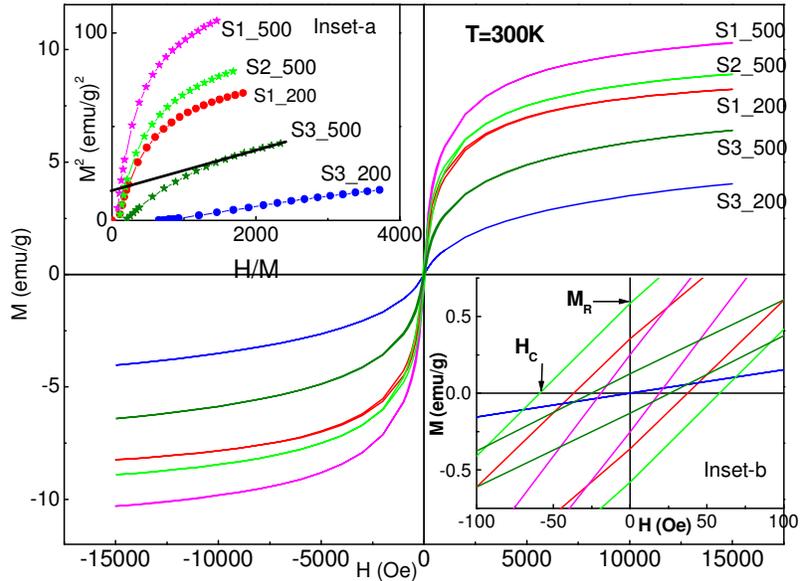

Fig. 5 (Color online) M(H) loops of composite samples at 300 K. Inset-a shows the Arrot plot for visualizing the spontaneous magnetization. Inset-b shows the low field loops which are used to calculate $M_R$ and $H_C$.

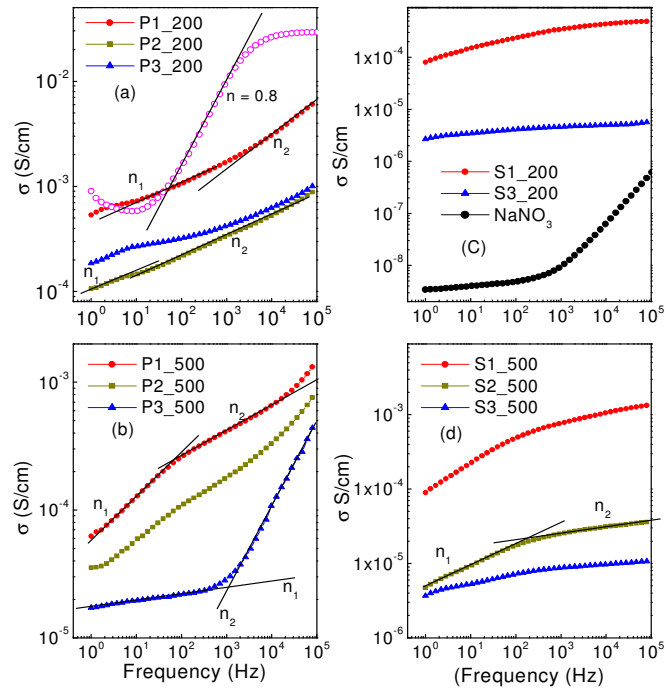

Fig. 6 (Colour online) Frequency dependent ac conductivity of different samples. The lines guide to the linear fit of the conductivity data in log-log scale according to Jonscher power law with exponents ($n_1$ and $n_2$).

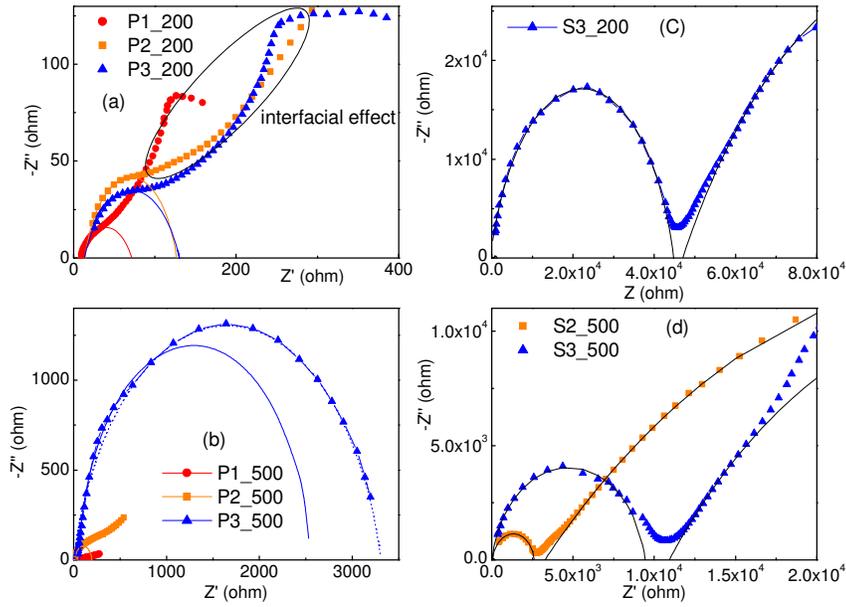

Fig. 7. (Color online) (log-log scale) Cole-Cole plots for different pH values Pure samples at 200 °C (a) and at 500 °C (b), composite samples at 200 °C (C) and at 500 °C (d). The lines are fit curves of data to semi-circles.

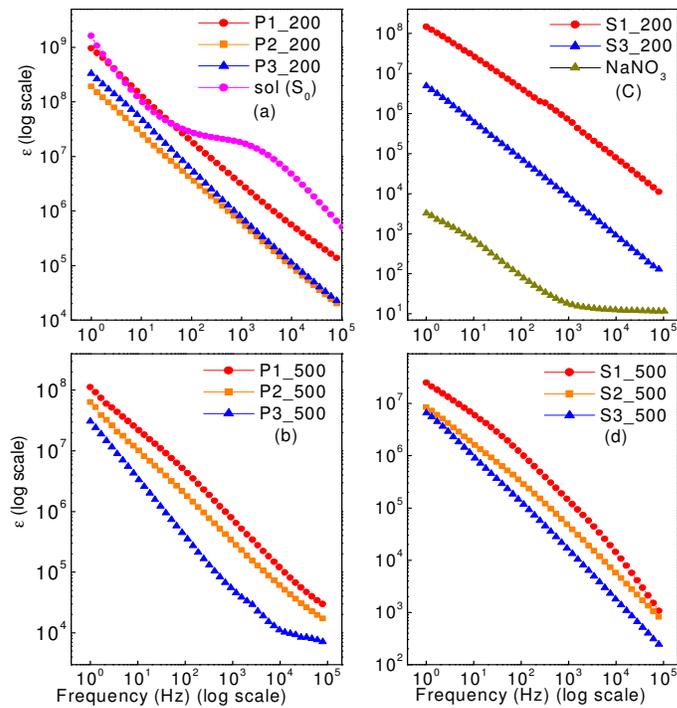

Fig. 8 (Colour online) Frequency dependent dielectric constant for sol ($S_0$) and pure samples at 200 °C (a), at 500 °C (b), and composite samples at 200 °C, at 500 °C (d).

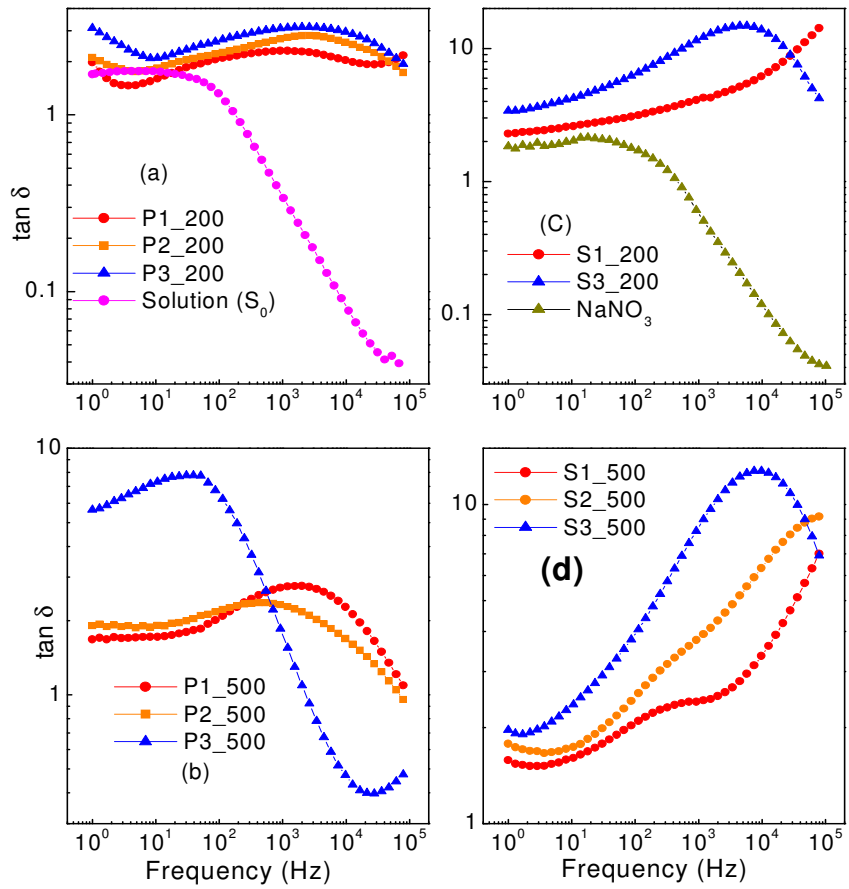

Fig. 9. (Color online) (log-log scale) Frequency dependent dielectric loss (tanδ) for stock solution and Pure samples at 200 $^0$C (a), Pure samples at 500 $^0$C (b), composite samples at 200 $^0$C (C) and composite samples at 500 $^0$C (d).

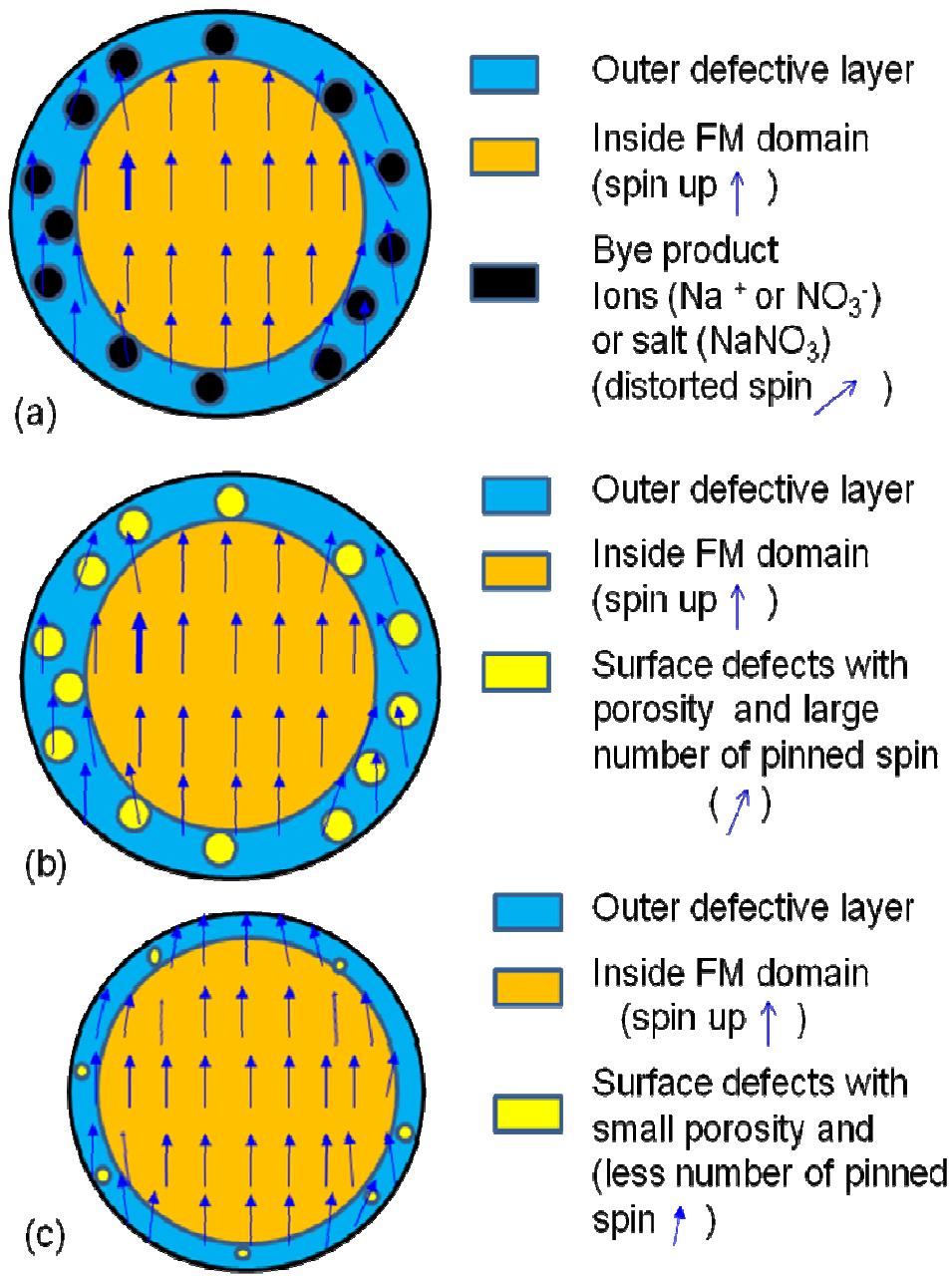

Fig. 10 (Colour online) Schematic diagram of the chemical routed ferrite particle at different stages after co-precipitation: (a) as-prepared after washing with distilled water, (b) after heating the as-prepared sample at lower temperature, say 200 °C in our case, (c) after heating the as-prepared sample at relatively higher temperature, say 500 °C in our case.